\documentclass[conference, letterpaper]{IEEEtran}

\usepackage{ifthen}
\usepackage[cmex10]{amsmath}
\usepackage{color}
\usepackage{subfigure}
\usepackage{amsmath}
\usepackage{mathtools}
\usepackage{amssymb}
\usepackage{marvosym}
\usepackage{type1cm}
\usepackage{epic}
\usepackage{wasysym}
\usepackage{latexsym}
\usepackage{graphicx}
\usepackage{placeins}
\usepackage{subfigure}
\usepackage{url}
\usepackage{fancyhdr}
\usepackage{wrapfig}
\usepackage{cite}
\usepackage{url}
\usepackage{psfrag}
\usepackage{enumerate}
\newcommand{\set}[1]{\mathcal{#1}}   
\newcommand{\wt}[1]{\widetilde{#1}}   

\newtheorem{theorem}{Theorem}

\newtheorem{remark}{Remark}
\newtheorem{lemma}{Lemma}

\begin{document}

\title{Informational Divergence Approximations\\ to Product Distributions}

\IEEEoverridecommandlockouts

 \author{
\IEEEauthorblockN{Jie Hou and Gerhard Kramer}
\IEEEauthorblockA{Institute for Communications Engineering\\
   Technische Universit\"at M\"unchen, 80290 Munich, Germany\\
       Email: \{jie.hou, gerhard.kramer\}@tum.de}
}

\maketitle

\begin{abstract}
The minimum rate needed to accurately approximate a product distribution based on an unnormalized informational divergence is
shown to be a mutual information. This result subsumes results of Wyner on common information and Han-Verd\'{u} on
resolvability. The result also extends to cases where the source distribution is unknown but the entropy is known.
\end{abstract}

\section{Introduction}
What is the minimal rate needed to generate a good approximation of a target distribution with respect to some distance measure?
For example, to learn a system response, we might give inputs to the system and compute the output statistics. However, in
computer simulations the inputs are only some approximations of the true distributions that are generated with random number
generators. We would like to use a small number of bits to generate good approximations of a target distribution.

Wyner considered such a problem and characterized the smallest rate needed to approximate a {\em
product} distribution accurately when using the {\em normalized} informational divergence as the distance measure between two
distributions. The smallest rate is a Shannon mutual information \cite{Wyner02}. Han-Verd\'{u} \cite{Verdu01} showed that the
same rate is necessary and sufficient to generate distributions arbitrarily close to an {\em information
stable} distribution in terms of {\em variational distance}. Note that normalized informational divergence and
variational distance are not necessarily larger or smaller than the other.


The main contributions of this work are to show that the minimal rate needed to make the {\em unnormalized} informational
divergence between a target product distribution and the approximating distribution arbitrarily small is the same Shannon mutual
information as in \cite{Wyner02, Verdu01} and we extend the proof to cases where the encoder has a
non-uniform input distribution. Our result implies results in \cite{Wyner02} and \cite{Verdu01} when restricting attention to
product distributions (in particular Theorem 6.3 in \cite{Wyner02} and Theorem 4 in \cite{Verdu01}). We remark that Hayashi
developed closely related theory via Gallager's error exponent in \cite{Ha01} and Bloch and Kliewer considered
non-uniform distributions for secrecy in \cite{Bloch02}. We also refer to results by Csiszar~\cite[p.~44, bottom]{Csiszar03} who 
treats strong secrecy by showing that a variational distance exhibits an exponential behavior with
block length $n$~\cite[Prop.~2]{Csiszar03}. This result implies that an unnormalized
mutual information expression can be made small with growing $n$ via~\cite[Lemma~1]{Csiszar03}.


The paper is organized as follows. In Section \ref{notation}, we state the problem. In Section \ref{Achievability} we state and
prove the main result. Section \ref{Diss} discusses related work and extensions.

\section{Preliminaries}
\label{notation}
Random variables are written with upper case letters and their realizations with the corresponding lower
case letters. Superscripts denote finite-length sequences of variables/symbols, e.g., $X^n=X_{1},\dots, X_{n}$. Subscripts
denote the position of a variable/symbol in a sequence. For instance, $X_{i}$ denotes the $i$-th variable in $X^n$. A random
variable $X$ has probability distribution $P_X$ and the support of $P_X$ is denoted as $\text{supp}(P_X)$. We write
probabilities with subscripts $P_{X}(x)$ but we drop the subscripts if the arguments of the distribution are lower case versions
of the random variables. For example, we write $P(x)=P_X(x)$. If the $X_i$, $i=1,\dots,n$, are independent and identically
distributed (i.i.d.) according to $P_X$, then we have $P(x^n)=\prod^n_{i=1}P_X(x_i)$ and we write $P_{X^n}=P^n_X$. Calligraphic
letters denote sets. The size of a set $\set S$ is denoted as $|\set S|$. We use $\set T^n_\epsilon(P_X)$ to denote the set
of
letter-typical sequences of length $n$ with respect to the probability distribution $P_X$ and the non-negative number $\epsilon$
\cite[Ch. 3]{ Massey01}, \cite{Roche01}, i.e., we have
\begin{align*}
\set T^n_\epsilon(P_X)=\left\{x^n:\Big| \frac{N(a|x^n)}{n} -P_X(a) \Big| \le \epsilon P_X(a),\; \forall a\in \set X  \right\}
\end{align*}
where $N(a|x^n)$ is the number of occurrences of $a$ in $x^n$.

\begin{figure}[t!]
\centering
\psfrag{W}[][][1]{$W=\{1,\dots, M\}$}
\psfrag{unW}[][][1]{$U^n$}
\psfrag{Vn}[][][1]{$V^n\sim P_{V^n}$}
\psfrag{Pvu}[][][1]{$Q^n_{V|U}$}
\psfrag{Pu}[][][1]{$Q_{U^n}$}
\psfrag{f}[][][1]{Encoder}
\includegraphics[width=7.5cm]{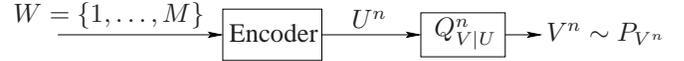}
\caption{Coding problem with the goal of making $P_{V^n} \approx Q^n_V$.}
\label{Approx}
\end{figure}

Consider the system depicted in Fig. \ref{Approx}. The random variable $W$ is {\em uniformly} distributed over $\{1,\dots,M\}$,
$M=2^{nR}$, and is encoded to sequences
\begin{align}
U^n=f(W).
\end{align}
$V^n$ is generated from $U^n$ through a memoryless channel $Q^n_{V|U}$ and has distribution $P_{V^n}$. A rate $R$ is {\em
achievable} if for any $\xi>0$ there is a sufficiently large $n$ and an encoder such that
\begin{align}
D(P_{V^n}||Q^n_V)=\sum_{v^n\in \text{supp}(P_{V^n})} P(v^n)\log\frac{P(v^n)}{Q^n_V(v^n)}\label{divergence}
\end{align}
is less than $\xi$. We wish to determine the smallest achievable rate.

\section{Main Result and Proof}
\label{Achievability}
 \begin{theorem}
For a given target distribution $Q_V$, the rate $R$ is achievable if $R>I(V;U)$, where $I(V;U)$ is calculated with some joint
distribution $Q_{UV}$ that has marginal $Q_V$ and $|\text{supp}(Q_U)|\le |\set V|$. The rate $R$ is not achievable if $R<I(V;U)$
for all $Q_{UV}$ with $|\text{supp}(Q_U)|\le |\set V|$.
\end{theorem}
We provide two proofs, one with Shannon's typicality argument and the other with Gallager's error exponent
\cite{Gallager01} where we extend results in \cite{Ha01}. Suppose $U$ and $V$ have finite alphabets $\set U$ and $\set V$,
respectively. Let $Q_{UV}$ be a probability distribution with marginals $Q_U$ and $Q_V$. Let $U^nV^n\sim Q^n_{UV}$, i.e., for any
$u^n\in \set U^n$, $v^n\in \set V^n$ we have
\begin{align}
Q(u^n, v^n)&=\prod^n_{i=1} Q_{UV}(u_i, v_i)=Q^n_{UV}(u^n, v^n)\\
Q(v^n|u^n)&=\prod^n_{i=1} Q_{V|U}(v_i|u_i)=Q^n_{V|U}(v^n|u^n).
\end{align}

\begin{figure}[t!]
\centering
\psfrag{W}[][][1]{$W$}
\psfrag{un}[][][1]{$U^n(\cdot)$}
\psfrag{unW}[][][1]{$U^n(W)$}
\psfrag{Qvu}[][][1]{$Q^n_{V|U}$}
\psfrag{Vn}[][][1]{$V^n$}
\psfrag{u1}[][][1]{$U^n(1)$}
\psfrag{u2}[][][1]{$U^n(2)$}
\psfrag{uM}[][][1]{$U^n(M)$}
\psfrag{dot}[][][1]{$\cdots$}
\psfrag{Qun}[][][1]{$Q^n_U$}
\psfrag{Encoder}[][][1]{Encoder}
\includegraphics[width=8.7cm]{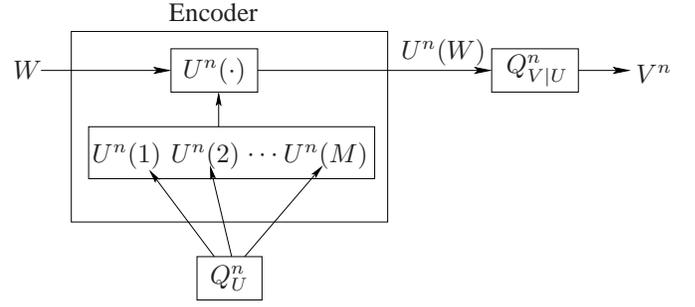}
\caption{The random coding experiment.}
\label{randomcoding}
\end{figure}

Let $\set C=\{U^n(w)\}^M_{w=1}$, where the $U^n(w), w=1,\dots, M$, are generated in an i.i.d. manner using $Q^n_U$. $V^n$
is generated from $U^n(W)$ through the channel $Q^n_{V|U}$ (see Fig. \ref{randomcoding}). We have
\begin{align}
P(v^n)=\sum^{M}_{w=1} \frac{1}{M} \cdot Q^n_{V|U}(v^n|u^n(w)). \label{actual}
\end{align}
Note that if for a $v^n$ we have
\begin{align}
Q^n_V(v^n)=\sum_{u^n \in \text{supp}(Q^n_U)} Q^n_U(u^n)Q^n_{V|U}(v^n|u^n)=0
\end{align}
then we have
\begin{align}
Q^n_{V|U}(v^n|u^n)=0,\;\text{for all}\; u^n\in \text{supp}(Q^n_U).
\end{align}
This means $P(v^n)=0$ and $\text{supp}(P_{V^n})
\subseteq \text{supp}(Q^n_{V}) $ so that $D(P_{V^n}||Q^n_V)< \infty$. We further have
\begin{align}
\text{E}\left[\frac{Q^n_{V|U}(v^n|U^n)}{Q^n_V(v^n)}\right]&=\sum_{u^n}Q^n_{U}(u^n)\cdot \frac{Q^n_{V|U}(v^n|u^n)}{Q^n_V(v^n)
}=1.
\label{condition}
\end{align}

\subsection{Shannon's Typicality}
The average informational divergence over $W$, $\set C$ and $V^n$ is (recall that $P(w)=\frac{1}{M}, w=1,\dots, M$):
\begin{align}
&\text{E}[D(P_{V^n}||Q^n_{V}) ]\overset{(a)}{=}\text{E}\left[ \log \frac{ \sum^M_{j=1} \frac{1}{M} \cdot
Q_{V^n|U^n}(V^n|U^n(j))}{Q^n_V(V^n)}\right]\notag\\
&=\sum^M_{w=1} \frac{1}{M} \cdot \text{E}\left[ \log \frac{ \sum^M_{j=1}
Q^n_{V|U}(V^n|U^n(j))}{MQ^n_V(V^n)}\Bigg|W=w\right]\notag\\
&\overset{(b)}{\le} \sum^M_{w=1} \frac{1}{M} \cdot \text{E}\left[ \log \left( \frac{ Q^n_{V|U}(V^n|U^n(w))}{MQ^n_V(V^n)} +
\frac{M-1}{M}\right)\Bigg|W=w\right]\notag\\
&\le \sum^M_{w=1} \frac{1}{M}\cdot\text{E}\left[ \log \left( \frac{ Q^n_{V|U}(V^n|U^n(w))}{MQ^n_V(V^n)}
+1\right)\Bigg|W=w\right]\notag\\
&\overset{(c)}{=}\text{E} \left[\log\left( \frac{ Q^n_{V|U}(V^n|U^n) }{M\cdot Q^n_V(V^n)} + 1\right)\right] \label{star}
\end{align}
where
\begin{enumerate}[(a)]
\item follows by taking the expectation over $W$, $V^n$ and $U^n(1),\dots, U^n(M)$;
\item follows by the concavity of the logarithm and Jensen's inequality applied to the expectation over the $U^n(j), j\ne w$, and
by using (\ref{condition});
\item follows by choosing $U^nV^n\sim Q^n_{UV}$.
\end{enumerate}
Alternatively, we can make the steps (\ref{star}) more explicit:
\begin{align}
&\text{E}[D(P_{V^n}||Q^n_{V}) ]\overset{(a)}{=} \sum_{u^n(1)}\cdots\sum_{u^n(M)}\prod^M_{k=1}{Q^n_{U}(u^n(k))}\notag\\
&\sum_{v^n}\sum^M_{w=1} \frac{1}{M}\cdot Q^n_{V|U}(v^n|u^n(w))\left[\log \frac{\sum^M_{j=1} Q^n_{V|U}(v^n|u^n(j))}{M\cdot
Q^n_{V}(v^n)}\right]\notag\\
&=\sum^M_{w=1}\frac{1}{M}\sum_{v^n}\sum_{u^n(w)}Q^n_{UV}(u^n(w), v^n)\notag\\
& {\sum^M_{k\ne w}} \sum_{u^n(k)} {\prod^M_{l\ne w}}{Q^n_{U}(u^n(l))}\left[\log
\frac{\sum^M_{j=1} Q^n_{V|U}(v^n|u^n(j))}{M\cdot
Q^n_{V}(v^n)}\right]\notag\\
&\overset{(b)}{\le}\sum^M_{w=1} \frac{1}{M} \sum_{v^n}\sum_{u^n(w)} Q^n_{UV}(u^n(w),v^n)\notag\\
&\left[\log\left( \frac{ Q^n_{V|U}(v^n|u^n(w)) }{M\cdot Q^n_{V}(v^n)}  +
\sum^M_{j\ne w} \sum_{u^n(j)}\left[\frac{Q^n_{UV}(u^n(j), v^n) }{M\cdot Q^n_{V}(v^n)}\right]\right)\right]\notag\\
&=\sum^M_{w=1}\frac{1}{M} \sum_{v^n}\sum_{u^n(w)}Q^n_{UV}(u^n(w), v^n)\notag\\
&\left[\log\left( \frac{ Q^n_{V|U}(v^n|u^n(w)) }{M\cdot Q^n_{V}(v^n)}  + \frac{M-1}{M}\right)\right]\notag\\
&\le \sum^M_{w=1} \frac{1}{M} \sum_{v^n}\sum_{u^n(w)}Q^n_{UV}(u^n(w), v^n)\notag\\
&\quad \left[\log\left(\frac{ Q^n_{V|U}(v^n|u^n(w)) }{M\cdot Q^n_V(v^n)}  + 1\right)\right]\notag\\
&\overset{(c)}{=}\text{E} \left[\log\left( \frac{ Q^n_{V|U}(V^n|U^n) }{M\cdot Q^n_V(V^n)}  + 1\right)\right].\label{star1}
\end{align}
We remark that the identity after $(a)$ is valid for $M=1$ by interpreting the empty sum followed by an empty product to be $1$.
We may write (\ref{star}) or (\ref{star1}) as
\begin{align}
\text{E} \left[\log\left(\frac{ Q^n_{V|U}(V^n|U^n) }{M\cdot Q^n_V(V^n)}  + 1\right)\right]=d_1+d_2 \label{average}
\end{align}
where
\begin{align*}
&d_1=\sum_{ (u^n, v^n )\in \set T^n_\epsilon(Q_{UV})}Q(u^n,v^n)
\log \left( \frac{Q(v^n|u^n)}{M\cdot Q(v^n)}+1 \right)\\
&d_2=\underset{(u^n, v^n) \in \text{supp}(Q^n_{UV}) } {\sum_{ (u^n, v^n )\notin \set T^n_\epsilon(Q_{UV})}} Q(u^n, v^n) \log
\left(
\frac{Q(v^n|u^n)}{M\cdot Q(v^n)}+1 \right).
\end{align*}
Using standard inequalities (see \cite{Roche01}) we have
\begin{align}
d_1&\le \sum_{ (u^n, v^n )\in \set T^n_\epsilon(Q_{UV})}Q(u^n,v^n) \log \left(
\frac{ 2^{-n(1-\epsilon)H(V|U)} }{M\cdot 2^{-n (1+\epsilon ) H(V)}}+1 \right)\notag\\
&\le \log \left( \frac{ 2^{-n(1-\epsilon)H(V|U)} }{M\cdot 2^{-n (1+\epsilon ) H(V)}}+1 \right)\notag\\
&=\log \left(2^{-n(R-I(V;U)-\epsilon(H(V|U)+H(V)) )} +1 \right)\notag\\
&\le \log(e)\cdot 2^{-n(R-I(V;U)-2\epsilon H(V))} \label{d1distance}
\end{align}
and $d_1\rightarrow 0$ if $R>I(V;U)+2\epsilon H(V)$ and $n\rightarrow
\infty$. We further have
\begin{align}
d_2&\le \underset{(u^n, v^n) \in \text{supp}(Q^n_{UV}) } {\sum_{ (u^n, v^n )\notin \set T^n_\epsilon(Q_{UV})}}Q(u^n, v^n)
\log
\left( \left(\frac{1}{\mu_V}\right)^n+1 \right)\notag\\
&\le 2|\set V|\cdot |\set U|\cdot e^{-2n\epsilon^2\mu^2_{UV}} \log \left (\left(\frac{1}{\mu_V}\right)^n+1
\right)\\
&\le 2|\set V|\cdot |\set U| \cdot e^{-2n\epsilon^2\mu^2_{UV}} \cdot n \cdot \log \left(\frac{ 1}{\mu_V}
+1\right) \label{d2distance}
\end{align}
and $d_2\rightarrow 0$ as $n\rightarrow \infty$, where
\begin{align}
\mu_{V}&=\text{min}_{ v \in \text{supp}(Q_{V})}Q(v)\\
\mu_{UV}&=\text{min}_{ (v, u)\in \text{supp}(Q_{UV})}Q(u,v).
\end{align}

Combining the above we have
\begin{align}
\text{E}[D(P_{V^n}||Q^n_{V}) ]\rightarrow 0 \label{goodcode}
\end{align}
if $R>I(V;U)+2\epsilon H(V)$ and $n\rightarrow \infty$. As usual, (\ref{goodcode}) means that there must exist a code with
$D(P_{V^n}||Q^n_{V})<\xi$ for any $\xi>0$ and sufficiently large $n$. This proves the coding theorem. The converse follows
from \cite[Theorem 5.2]{Wyner02} by removing the normalization factor $\frac{1}{n}$.

\begin{remark}
\label{remark2}
The cardinality bound on $\text{supp}(Q_U)$. can be derived using techniques from \cite[Ch. 15]{Csiszar01}.
\end{remark}

\begin{remark}
\label{remark3}
If $V=U$, then we have $R>H(V)$.
\end{remark}

Theorem 1 is proved using a uniform $W$ which represents strings of uniform bits. If we use a non-uniform $W$ for
the coding scheme, can we still drive the unnormalized informational divergence to zero? We give the answer in
the following lemma.
\begin{lemma}
\label{lem:lemma1}
Let $W=B^{nR}$ be a bit stream with $nR$ bits that are generated i.i.d. with a binary distribution $P_X$ with $P_X(0)=p$,
$0< p \le \frac{1}{2}$. The rate $R$ is achievable if
\begin{align}
R>\frac{I(V;U)}{H_2(p)}
\end{align}
where $H_2(\cdot)$ is the binary entropy function.
\end{lemma}
\begin{IEEEproof}
The proof is given in Appendix \ref{nonuniform}.
\end{IEEEproof}
\begin{remark}
Lemma~1 states that even if $W$ is not uniformly distributed, the informational divergence can
be made small. This is useful because if the distribution of $W$ is not known exactly, then we can choose $R$ large enough to
guarantee the desired resolvability result. A similar result was developed in \cite{Bloch02} for secrecy.

\end{remark}

\subsection{Gallager's Error Exponent}
\label{galla}
We provide a second proof using Gallager's error exponent \cite{Gallager01} by extending \cite[Lemma 2]{Ha01} to asymptotic
cases. Consider $-\frac{1}{2}\le \rho \le 0$ and define
\begin{align}
E^n_0(\rho, Q^n_{UV})&=\log_2 \sum_{v^n}\left\{ \text{E}[P(v^n)^\frac{1}{1+\rho}]  \right\}^{1+\rho}\\
E_0(\rho, Q_{UV})&=\log_2 \sum_{v}\left\{ \sum_u Q(u) Q(v|u)^\frac{1}{1+\rho}  \right\}^{1+\rho}\\
E_G(R, Q_{UV})&=\inf_{-\frac{1}{2}\le \rho<0} \;\left\{E_0(\rho, Q_{UV})+\rho R\right\}.
\end{align}
Due to \cite[Lemma 2]{Ha01}, we have the following properties concerning $E^n_0(\rho, Q^n_{UV})$ and $E_0(\rho,
Q_{UV})$:

{\em Property 1:}
\begin{align}
E^n_0(0, Q^n_{UV})&=E_0(0, Q_{UV})=0
\end{align}

{\em Property 2:}
\begin{align}
\left. \frac{\partial E^n_0(\rho, Q^n_{UV})}{\partial \rho} \right|_{\rho=0}&=-\text{E}[D(P_{V^n}||Q^n_V)]\notag\\
\left. \frac{\partial E_0(\rho, Q_{UV})}{\partial \rho} \right|_{\rho=0}&=-I(V;U)
\end{align}

{\em Property 3:}
\begin{align}
\frac{\partial^2 E^n_0(\rho, Q^n_{UV})}{\partial \rho^2}&\ge 0\notag\\
\frac{\partial^2 E_0(\rho, Q_{UV})}{\partial \rho^2}&\ge 0
\end{align}

Due to \cite[Theorem 5.6.3]{Gallager01}, we have
\begin{align}
\label{GaOpt}
\left\{
\begin{array}{ll}
E_G(R, Q_{UV})<0 &  \text{if}\; R>I(V;U)\\
E_G(R, Q_{UV})=0 &  \text{if}\; R\le I(V;U)
\end{array}
\right.
\end{align}

By extending \cite[Sec. III, Inequality (15)]{Ha01} to asymptotic cases, we have the following lemma.
\begin{lemma}
\label{lem2}
We have
\begin{align}
E^n_0(\rho, Q^n_{UV})\le \log_2\left(1+2^{nE_G(R, Q_{UV})}\right).
\end{align}
\end{lemma}
\begin{IEEEproof}
The proof is given in Appendix \ref{GE}.
\begin{figure}[t!]
\centering
\psfrag{f1}[][][1]{$E^n_0(\rho, Q^n_{UV})$}
\psfrag{slope1}[][][1]{$\rho\cdot (-\text{E}[D(P_{V^n}||Q^n_V)] )$}
\psfrag{f2}[][][1]{$E_0(\rho, Q_{UV})$}
\psfrag{slope2}[][][1]{$\rho\cdot (-I(V;U) )$}
\psfrag{0}[][][1]{$0$}
\psfrag{rho}[][][1]{$\rho$}
\psfrag{half}[][][1]{$-\frac{1}{2}$}
\includegraphics[width=7.7cm]{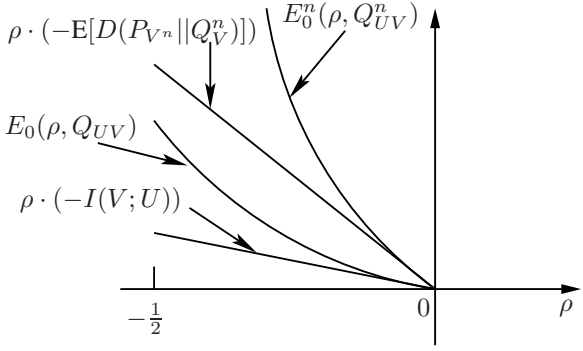}
\caption{An example of $E^n_0(\rho, Q^n_{UV})$ and $E_0(\rho, Q_{UV})$.}
\label{illu}
\end{figure}
\end{IEEEproof}
Combining Properties $1$-$3$, we have $E^n_0(\rho, Q^n_{UV})$ and $E_0(\rho, Q_{UV})$ are convex in $\rho$, for $-\frac{1}{2}\le
\rho\le 0$ and (see Fig.~\ref{illu})
\begin{align}
\rho\cdot (-\text{E}[D(P_{V^n}||Q^n_V)])&\le E^n_0(\rho, Q^n_{UV})
\end{align}
which means
\begin{align}
\label{GaInequ}
\text{E}[D(P_{V^n}||Q^n_V)]&\le \frac{E^n_0(\rho, Q^n_{UV})}{-\rho}\notag\\
&\overset{(a)}{\le} \frac{\log_2 \left(1+2^{nE_G(R, Q_{UV})}\right)}{-\rho}
\end{align}
where $(a)$ follows from Lemma~\ref{lem2}. The right hand side of (\ref{GaInequ}) goes to $0$ as $n\rightarrow \infty$ as long as
(see (\ref{GaOpt}))
\begin{align}
R>I(V;U).
\end{align}

\begin{remark}
\label{remark1}
This proof applies to {\em continuous} random variables by replacing the sums in the proof of
Lemma~\ref{lem2} with integrals.
\end{remark}

\begin{remark}
The average divergence $\text{E}[D(P_{V^n}||Q^n_V)]$ can be viewed as the mutual information $I(\set C; V^n)$ from
the random codebook $\set C$ to the output $V^n$ \cite[Sec. III]{Ha01}. To show this, denote $\set {\wt C}$ as a realization
of $\set C$ and we have (see (\ref{star1}))
\begin{align}
&I(\set C;V^n)=\sum_{\set {\wt C}}P(\set {\wt C})\sum_{v^n}P(v^n|\set {\wt C})\log \frac{P(v^n|\set {\wt
C})}{Q^n_V(v^n)}\notag\\
&=\sum_{u^n(1)}\cdots \sum_{u^n(M)} \prod^M_{k=1} Q^n_{U}(u^n(k))\notag\\
&\sum_{v^n} \sum^M_{w=1} \frac{1}{M}  \cdot Q^n_{V|U}(v^n|u^n(w))\log \frac{ \sum^M_{j=1} \frac{1}{M}
Q^n_{V|U}(v^n|u^n(j))}{Q^n_V(v^n)}\notag\\
&=\text{E}\left [\log \frac{ \sum^M_{j=1} \frac{1}{M} Q^n_{V|U}(V^n|U^n(j))}{Q^n_V(V^n)}\right]\notag\\
&=\text{E}[D(P_{V^n}||Q^n_V)].
\end{align}
Thus, as $\text{E}[D(P_{V^n}||Q^n_V)]\rightarrow 0$ we have $I(\set C; V^n)\rightarrow 0$ which means that $\set C$ and
$V^n$ are (almost) independent. This makes sense, since as $P_{V^n}\rightarrow Q^n_V$ one is not able to
distinguish which codebook is used to generate the output.
\end{remark}

\section{Discussion}
\label{Diss}
Hayashi studied  the resolvability problem using unnormalized divergence and he derived bounds for nonasymptotic cases
\cite[Lemma 2]{Ha01}. We have outlined his proof steps in Sec. \ref{galla}. Theorem~1 can be derived by extending \cite[Lemma
2]{Ha01} to asymptotic cases (see \ref{galla}) and it seems that such a result was the underlying motivation for \cite[Lemma
2]{Ha01}. Unfortunately, Theorem 1 is not stated explicitly in \cite{Ha01} and the ensuing asymptotic analysis was done for {\em
normalized} informational divergence. Hayashi's proofs (he developed two approaches) were based on Shannon random coding.

Theorem 1 implies \cite[Theorem 6.3]{Wyner02} which states that for $R>I(V;U)$ the normalized divergence
$\frac{1}{n}D(P_{V^n}||Q^n_V)$ can be made small. Theorem 1 implies \cite[Theorem 4]{Verdu01} for product distributions through
Pinsker's inequality \cite[Lemma 11.6.1]{Cover01}
\begin{align}
D(P_X||Q_X)\ge \frac{1}{2 \ln 2} ||P_X-Q_X||^2_\text{TV}
\end{align}
where
\begin{align}
||P_X-Q_X||_\text{TV}=\sum_{x}|P(x)-Q(x)|.
\end{align}
Moreover, the speed of decay in  (\ref{d1distance}) and (\ref{d2distance}) is (almost) exponential with $n$. We can thus make
\begin{align}
\alpha(n) \cdot \text{E}\left[D(P_{V^n}||Q^n_V)\right]
\end{align}
vanishingly small as $n\rightarrow \infty$, where $\alpha(n)$ represents a {\em sub-exponential} function of $n$ that satisfies,
\begin{align}
\lim_{n\rightarrow \infty} \frac{n\cdot \alpha(n)}{e^{\beta n}}=0
\end{align}
where $\beta$ is positive and independent of $n$ (see also \cite{Ha01}). For example, we may choose $\alpha(n)=n^m$ for any
integer $m$. We may also choose $\alpha(n)=e^{\gamma n}$ where $\gamma < \beta$.



Since all achievability results in \cite{Cuff01} are based on \cite[Theorem 4]{Verdu01}, Theorem 1 extends the results in
\cite{Cuff01} as well. Theorem 1 is closely related to {\em strong} secrecy \cite{Bloch01} and provides a simple proof
that Shannon random coding suffices to drive an {\em unnormalized} mutual information between messages and eavesdropper
observations to zero.

Theorem 1 is valid for approximating product distributions only. However extensions to a broader class of
distributions, e.g., {\em information stable} distributions \cite{Verdu01}, are clearly possible.

Finally, an example code is as follows (courtesy of F. Kschischang). Consider a channel with input and output alphabet the $2^7$
binary $7$-tuples. Suppose the channel maps each input uniformly to a $7$-tuple that is distance $0$ or $1$ away, i.e., there are
$8$ channel transitions for every input and each transition has probability $\frac{1}{8}$. A simple \lq\lq modulation\rq\rq\;code
for this channel is the $(7,4)$ Hamming code. The code is perfect and if we choose each codeword with probability $\frac{1}{16}$,
then the output $V^7$ of the channel is uniformly distributed over all $2^7$ values. Hence $I(V;U)=4$ bits suffice to
\lq\lq approximate\rq\rq\; the product distribution (here there is no approximation).


\begin{appendices}
\label{App}

\section{Non-Uniform $W$}
\label{nonuniform}
Observe that $H(W)=H(B^{nR})=nR\cdot H_2(p)$. Following the same steps as in (\ref{star}) we have
\begin{align}
&\text{E}[D(P_{V^n}||Q^n_{V}) ]=\text{E}\left[ \log \frac{ \sum^M_{j=1}
P(j) Q_{V^n|U^n}(V^n|U^n(j))}{Q^n_V(V^n)}\right]\notag\\
&=\sum_{w} P(w)\cdot  \text{E}\left[ \log \frac{ \sum^M_{j=1} P(j) Q^n_{V|U}(V^n|U^n(j))}{Q^n_V(V^n)}\Bigg|W=w\right]\notag\\
&\le \sum_{w}P(w)\cdot \text{E}\left[ \log \left( \frac{P(w)Q^n_{V|U}(V^n|U^n(w))}{Q^n_V(V^n)}
+1-P(w)\right)\right]\notag\\
&\le \sum_{w}P(w)\cdot \text{E}\left[ \log \left( \frac{P(w) Q^n_{V|U}(V^n|U^n(w))}{Q^n_V(V^n)}
+1\right)\right]\notag\\
&=d_1+d_2 +d_3
\end{align}
where
\begin{align}
d_1&=\sum_{w\in \set T^n_\epsilon(P^n_X)}P(w)\sum_{ (u^n(w),v^n)\in \set
T^n_\epsilon(Q^n_{UV})}Q^n_{UV}(u^n(w),v^n)\notag\\
&\quad\quad\left[ \log \left( \frac{P(w) Q^n_{V|U}(v^n|u^n(w))}{Q^n_V(v^n)} +1\right)\right]\notag\\
d_2&=\sum_{w\in \set  T^n_\epsilon(P^n_X)}P(w)    \underset{(u^n(w), v^n) \in \text{supp}(Q^n_{UV}) } {\sum_{ (u^n(w), v^n
)\notin \set T^n_\epsilon(Q_{UV})}} Q^n_{UV}(u^n(w),v^n)\notag\\
&\quad\quad\left[ \log \left( \frac{P(w) Q^n_{V|U}(v^n|u^n(w))}{Q^n_V(v^n)} +1\right)\right]\notag\\
d_3&=\underset{w\in \text{supp}(P^n_X)}{\sum_{w\notin \set T^n_\epsilon(P^n_X)}}P(w)\sum_{ (u^n(w),v^n)
\in \text{supp} (Q^n_{UV})}Q^n_{UV}(u^n(w),v^n)\notag\\
&\quad\quad\left[ \log \left( \frac{P(w) Q^n_{V|U}(v^n|u^n(w))}{Q^n_V(v^n)} +1\right)\right].
\end{align}

We can bound $d_1$ as follows (see (\ref{d1distance}))
\begin{align}
d_1&\le \sum_{w\in \set T^n_\epsilon(P^n_X)}P(w) \left[ \log \left( \frac{2^{n(I(V;U) +2\epsilon H(V))}
}{2^{n(1-\epsilon)R\cdot H_2(p) } } +1\right)\right]\notag\\
&\le  \log \left( 2^{-n\left(R\cdot H_2(p)- I(V;U) -\epsilon(2H(V)+R\cdot H_2(p))\right)}   +1\right)\notag\\
&\le \log(e)\cdot 2^{-n\left(R\cdot H_2(p)- I(V;U) -\delta_\epsilon(n)\right)}
\end{align}
which goes to zero if $R>\frac{I(V;U)+\delta_\epsilon(n)}{H_2(p)}$ and $n\rightarrow \infty$, where
$\delta_\epsilon(n)=\epsilon(2H(V)+R\cdot H_2(p))$. We also have
\begin{align}
d_2&\le \sum_{w\in \set T^n_\epsilon(P^n_X)}P(w)  \underset{(u^n(w), v^n) \in \text{supp}(Q^n_{UV}) } {\sum_{ (u^n(w), v^n
)\notin \set T^n_\epsilon(Q_{UV})}} Q^n_{UV}(u^n(w),v^n)\notag\\
&\quad\quad\left[ \log \left( \left(\frac{1}{\mu_V}\right)^n +1\right)\right]\notag\\
&\le 2|\set V|\cdot |\set U|\cdot e^{-2n\epsilon^2\mu^2_{UV}}  \log \left(\left(\frac{1}{\mu_V}\right)^n
+1\right)
\end{align}
which goes to zero as $n\rightarrow \infty$ (see (\ref{d2distance})). We further have
\begin{align}
d_3&\le \underset{w\in \text{supp}(P^n_X)}{\sum_{w\notin \set T^n_\epsilon(P^n_X)}}P(w)\sum_{ (u^n(w),v^n)\in
\text{supp}(Q^n_{UV})} Q^n_{UV}(u^n(w),v^n)\notag\\
&\quad\quad\left[ \log \left( \left(\frac{1}{\mu_V}\right)^n +1\right)\right]\notag\\
&\le \underset{w\in \text{supp}(P^n_X)} {\sum_{w\notin \set T^n_\epsilon(P^n_X)}}P(w)\left[ \log \left(
\left(\frac{1}{\mu_V}\right)^n
+1\right)\right]\notag\\
&\le 4\cdot e^{-2n\epsilon^2p^2}  \log \left(\left(\frac{1}{\mu_V}\right)^n+1\right)
\end{align}
which goes to zero as $n\rightarrow \infty$ (see (\ref{d2distance})).

Combining the above for non-uniform $W$ we have
\begin{align}
\text{E}[D(P_{V^n}||Q^n_V)] \rightarrow 0
\end{align}
if $R>\frac{I(V;U)+\delta_n(\epsilon)}{H_2(p)}$ and $n\rightarrow \infty$.

\section{Proof of Lemma 2}
\label{GE}
We extend the proof of \cite[Sec. III, Inequality (15)]{Ha01} to asymptotic cases to establish Lemma 2. Recall that
$-\frac{1}{2}\le \rho \le 0$. Let $s=\frac{-\rho}{1+\rho}$ so we have
\begin{align}
 &0\le s \le 1\notag\\
&1+s  = \frac{1}{1+\rho} \label{relation}
\end{align}
We also have for any $a, b\ge 0$ and $0 \le x \le 1$
\begin{align}
(a+b)^x\le a^x + b^x.\label{powerIneq}
\end{align}
Observe that for any $v^n$ we have
\begin{align}
\text{E}[P(v^n)]&=\text{E}\left[\sum^{M}_{w=1} \frac{1}{M}\cdot  Q^n_{V|U}(v^n|U^n(w))\right]\notag\\
&=\text{E}\left[Q^n_{V|U}(v^n|U^n(1))\right]\notag\\
&=\text{E}\left[\prod^n_{i=1} Q_{V|U}(v_i|U_i(1))\right]\notag\\
&=\prod^n_{i=1} \text{E}\left[ Q_{V|U}(v_i|U_i(1))\right]\notag\\
&=\prod^n_{i=1} \left[ \sum_{u} Q(u)Q_{V|U}(v_i|u)\right]\notag\\
&=\prod^n_{i=1} Q_V(v_i) =Q^n_V(v^n) \label{average1}
\end{align}

We further have
\begin{align}
&2^{E^n_0(\rho,Q^n_{UV})}=\sum_{v^n}\left\{ \text{E}[P(v^n)^\frac{1}{1+\rho}]  \right\}^{1+\rho}\notag\\
&\overset{(a)}{=}\sum_{v^n}\left\{ \text{E}[P(v^n)^{1+s}]  \right\}^\frac{1}{1+s}\notag\\
&=\sum_{v^n}\left\{ \text{E}\left[ \left( \sum^M_{w=1}\frac{1}{M} \cdot Q^n_{V|U}(v^n|U^n(w))\right)^{1+s}\right]
\right\}^\frac{1}{1+s}\notag\\
&=\frac{1}{M} \sum_{v^n}\left\{ \text{E}\left[ \sum^M_{w=1} Q^n_{V|U}(v^n|U^n(w)) \right.\right.\notag\\
&\left.\left.\left( Q^n_{V|U}(v^n|U^n(w))+ \sum^M_{j\ne w} Q^n_{V|U}(v^n|U^n(j)) \right)^{s} \right] \right\}^\frac{1}{1+s}
\label{441}
\end{align}
where $(a)$ follows from (\ref{relation}). Applying (\ref{powerIneq}) to (\ref{441}) we have
\begin{align}
&2^{E^n_0(\rho,Q^n_{UV})}\le \frac{1}{M} \sum_{v^n}\left\{ \text{E}\left[ \sum^M_{w=1} Q^n_{V|U}(v^n|U^n(w))
\right.\right.\notag\\
&\left.\left.\left(\left( Q^n_{V|U}(v^n|U^n(w))\right)^s+ \left(\sum^M_{j\ne w} Q^n_{V|U}(v^n|U^n(j)) \right)^{s}\right) \right]
\right\}^\frac{1}{1+s}\notag\\
&\overset{(a)}{=} \frac{1}{M} \sum_{v^n}\left\{ \text{E}\left[ \sum^M_{w=1} \left( Q^n_{V|U}(v^n|U^n(w))\right)^{1+s}\right]
\right.\notag\\ &\quad\quad\quad\quad\quad\quad +\sum^M_{w=1}\left(\text{E} \left [Q^n_{V|U}(v^n|U^n(w))\right] \right)\notag\\
&\left.\quad\quad\quad\quad\quad\quad\cdot \text{E} \left[\left(\sum^M_{j\ne w} Q^n_{V|U}(v^n|U^n(j) )
\right)^{s}\right]  \right\}^\frac{1}{1+s}\notag\\
&\overset{(b)}{\le} \frac{1}{M} \sum_{v^n}\left\{ M \text{E}\left[ \left( Q^n_{V|U}(v^n|U^n)\right)^{1+s}\right]
\right.\notag\\
&\left.+MQ^n_{V}(v^n) \cdot \left(\text{E} \left[\sum^M_{j\ne w} Q^n_{V|U}(v^n|U^n(j))\right]
\right)^{s}  \right\}^\frac{1}{1+s}\notag\\
&\overset{(c)}{=}\frac{1}{M} \sum_{v^n}\left\{ M \text{E}\left[ \left(
Q^n_{V|U}(v^n|U^n)\right)^{1+s}\right]\right.\notag\\
&\left.\quad\quad + MQ^n_{V}(v^n) \left((M-1)Q^n_{V}(v^n)\right)^{s}  \right\}^\frac{1}{1+s}\notag\\
&\le \frac{1}{M} \sum_{v^n}\left\{ M \text{E}\left[ \left( Q^n_{V|U}(v^n|U^n)\right)^{1+s}\right]+\left(MQ^n_{V}(v^n)\right)^{1+s}
 \right\}^\frac{1}{1+s}\label{442}
\end{align}
where
\begin{enumerate}[(a)]
\item follows because $U^n(w)$ is independent of $U^n(j)$, $j\ne w$
\item follows by choosing $U^nV^n\sim Q^n_{UV}$, by the concavity of $x^a$ for $0\le a\le 1$ and by (\ref{average1})
\item follows by (\ref{average1})
\end{enumerate}
Applying (\ref{powerIneq}) again to (\ref{442}) we have
\begin{align}
&2^{E^n_0(\rho,Q^n_{UV})}\le \frac{1}{M} \sum_{v^n}\left\{ \left(M \text{E}\left[
\left( Q^n_{V|U}(v^n|U^n)\right)^{1+s} \right]\right)^\frac{1}{1+s}\right.\notag\\
&\left.\quad\quad\quad\quad\quad\quad\quad\quad\quad\quad+MQ^n_{V}(v^n) \right\}\notag\\
&\overset{(a)}{=}1+ M^{\rho }\sum_{v^n} \left(\text{E}\left[
\left( Q^n_{V|U}(v^n|U^n)\right)^\frac{1}{1+\rho} \right]\right)^{1+\rho}\notag\\
&=1+ M^{\rho }\sum_{v^n} \left(\sum_{u^n}  Q^n_{U}(u^n)\left(
Q^n_{V|U}(v^n|u^n)\right)^\frac{1}{1+\rho}\right)^{1+\rho}\notag\\
&\overset{(b)}{=}1+ 2^{n\rho R }\sum_{v} \left(\sum_{u}  Q(u)\left(Q(v|u)\right)^\frac{1}{1+\rho}\right)^{n(1+\rho)}\notag\\
&=1+2^{n(E_0(\rho,Q_{UV}) +\rho R )}
\end{align}
where
\begin{enumerate}[(a)]
\item follows from (\ref{relation})
 \item follows because the $U_iV_i$ are i.i.d., $i=1,\dots,n$
\end{enumerate}
Optimizing over $\rho$, we have
\begin{align}
E^n_0(\rho,Q^n_{UV})\le \log_2\left(1+2^{nE_G(R,Q_{UV})}\right).
\end{align}

\end{appendices}

\section*{Acknowledgment}
J.~Hou and G.~Kramer were supported by an Alexander von Humboldt Professorship endowed by the
German Federal Ministry of Education and Research. G. Kramer was also supported by NSF Grant CCF-09-05235. The authors thank
G.~B\"{o}cherer, F.~Kschischang and M.~Bloch for useful remarks.

\bibliographystyle{IEEEtran}

\bibliography{bibfile}

\begin{thebibliography}{10}
\providecommand{\url}[1]{#1}
\csname url@samestyle\endcsname
\providecommand{\newblock}{\relax}
\providecommand{\bibinfo}[2]{#2}
\providecommand{\BIBentrySTDinterwordspacing}{\spaceskip=0pt\relax}
\providecommand{\BIBentryALTinterwordstretchfactor}{4}
\providecommand{\BIBentryALTinterwordspacing}{\spaceskip=\fontdimen2\font plus
\BIBentryALTinterwordstretchfactor\fontdimen3\font minus
  \fontdimen4\font\relax}
\providecommand{\BIBforeignlanguage}[2]{{%
\expandafter\ifx\csname l@#1\endcsname\relax
\typeout{** WARNING: IEEEtran.bst: No hyphenation pattern has been}%
\typeout{** loaded for the language `#1'. Using the pattern for}%
\typeout{** the default language instead.}%
\else
\language=\csname l@#1\endcsname
\fi
#2}}
\providecommand{\BIBdecl}{\relax}
\BIBdecl

\bibitem{Wyner02}
A.~Wyner, ``The common information of two dependent random variables,''
  \emph{IEEE Trans. Inf. Theory}, vol.~21, no.~2, pp. 163--179, March 1975.

\bibitem{Verdu01}
T.~Han and S.~Verd\'{u}, ``Approximation theory of output statistics,''
  \emph{IEEE Trans. Inf. Theory}, vol.~39, no.~3, pp. 752--772, May 1993.

\bibitem{Ha01}
M.~Hayashi, ``General nonasymptotic and asymptotic formulas in channel
  resolvability and identification capacity and their application to the
  wiretap channel,'' \emph{IEEE Trans. Inf. Theory}, vol.~52, no.~4, pp.
  1562--1575, April 2006.

\bibitem{Bloch02}
M.~Bloch and J.~Kliewer, ``On secure communication with constrained
  randomization,'' in \emph{IEEE Int. Symp. Inf. Theory}, Boston, MA, USA,
  2012, pp. 1172--1176.

\bibitem{Csiszar03}
I.~Csisz{\'a}r, ``Almost independence and secrecy capacity,'' \emph{Prob. of
  Inf. Transmission}, vol.~32, no.~1, pp. 40--47, Jan.--March 1996.

\bibitem{Massey01}
J.~L. Massey, \emph{Applied Digital Information Theory}, ETH Zurich, Zurich,
  Switzerland, 1980-1998.

\bibitem{Roche01}
A.~Orlitsky and J.~Roche, ``Coding for computing,'' \emph{IEEE Trans. Inf.
  Theory}, vol.~47, no.~3, pp. 903--917, March 2001.

\bibitem{Gallager01}
R.~G. Gallager, \emph{{Information Theory and Reliable Communication}}.\hskip
  1em plus 0.5em minus 0.4em\relax Wiley, 1968.

\bibitem{Csiszar01}
I.~Csisz{\'a}r and J.~K{\"o}rner, \emph{{Information Theory: Coding Theorems
  for Discrete Memoryless Systems}}.\hskip 1em plus 0.5em minus 0.4em\relax New
  York: Academic, 1981.

\bibitem{Cover01}
T.~Cover and J.~Thomas, \emph{Elements of Information Theory}, 2nd~ed.\hskip
  1em plus 0.5em minus 0.4em\relax New York: Wiley, 2006.

\bibitem{Cuff01}
P.~Cuff, H.~Permuter, and T.~Cover, ``Coordination capacity,'' \emph{IEEE
  Trans. Inf. Theory}, vol.~56, no.~9, pp. 4181--4206, Sept. 2010.

\bibitem{Bloch01}
M.~Bloch and J.~Barros, \emph{Physical Layer Security From Information Theory
  to Security Engineering}.\hskip 1em plus 0.5em minus 0.4em\relax Cambridge
  University Press, 2011.

\end{thebibliography}

\end{document}